\documentstyle[12pt,psfig,draft,lscape]{nature}
\begin{document}
\def\simlt{\mathrel{\hbox{\rlap{\hbox{\lower4pt\hbox{$\sim$}}}\hbox{$<$}}}}
\def\simgt{\mathrel{\hbox{\rlap{\hbox{\lower4pt\hbox{$\sim$}}}\hbox{$>$}}}}
\def\kms{${\rm km\;s^{-1}}$}
\def\mps{${\rm m\;s^{-1}}$}
%\psdraft

\title{\Large \bf A new transiting extrasolar giant planet}

\author
{Maciej Konacki\affiliation[1]
{Div.\ of Geological \& Planetary Sciences 150-21, California
Institute of Technology, Pasadena, CA 91125, USA},
Guillermo Torres\affiliation[2]
{Harvard-Smithsonian Center for Astrophysics, 60 Garden St.,
Cambridge, MA 02138, USA},
Saurabh Jha\affiliationmark[2]\affiliation[3]
{Department of Astronomy, University of California, Berkeley, CA
94720, USA} \&
Dimitar D.\ Sasselov\affiliationmark[2]
}

\dates{27 November 2002}{30 December 2002}
\headertitle{New Transiting Planet}
\mainauthor{Konacki, Torres, Jha, \&\ Sasselov}

\summary{ A conceptually simple and technologically feasible approach
to finding planets orbiting other stars is to observe the periodic
dimming of starlight due to a planet transiting in front of its star.
Despite many intense photometric searches, no transiting planet had
yet been discovered in this way. The only known transiting extrasolar
planet\cite{Char:00::,Hen:00::}, HD~209458b, was first detected by
precise radial velocity measurements\cite{Hen:00::,Mazeh:00::}. We
have measured radial velocities of a star, OGLE-TR-56, which shows a
1.2-day transit-like light curve found photometrically by Udalski et
al.\cite{U1:02::,U2:02::} Here we show that the velocity changes we
detect are probably induced by an object of 0.9 Jupiter masses --- a
very close-in gas-giant planet only 0.023~AU from its star, with a
planetary radius of 1.3 Jupiter radii and a mean density of
$\sim$0.5~g~cm$^{-3}$. At its small orbital distance, OGLE-TR-56b is
hotter than any known planet, approaching 1900~K, but it is stable
against long-term evaporation or tidal disruption. As the planet with
the tightest known orbit, OGLE-TR-56b will place strong constraints on
planet formation and migration models.}

\maketitle

\noindent The advent of high-precision Doppler and timing techniques
in the past decade has brought a rich bounty of giant
planets\cite{May:95::,Mar:98::,Schneider:02::} as well as smaller,
terrestrial-mass pulsar planets\cite{Wol:92::}. To date over one
hundred extrasolar giant planets have been found by different groups
using precise radial velocity measurements\cite{Schneider:02::}.
Photometric observations of transiting planets, when combined with
radial velocities, yield entirely new diagnostics: the planet size and
mean density\cite{Char:00::,Hen:00::}. Transits supply the orbital
inclination and a precise mass for the planet, and they additionally
enable a number of follow-up studies
\cite{Jha:00::,Brown:01::,Charbonneau:02::,Brown:02::}.  Hence, a
large number of transit searches are already underway or under
development\cite{Horne:03::}.  However, photometry alone cannot
distinguish whether the occulting object is a gas giant planet
($\sim$1--13 Jupiter masses), a brown dwarf ($\sim$13--80 Jupiter
masses) or a very late type dwarf star, because such objects have
nearly constant radius over a range from $\sim$0.001 to 0.1 Solar
masses. This critical parameter, the mass of the companion, can be
determined from the amplitude of the radial velocity variation induced
in the star.

One of the most successful searches to date is the Optical
Gravitational Lensing Experiment (OGLE), which uncovered 59 transiting
candidates in three fields in the direction of the Galactic centre
(OGLE-III)\cite{U1:02::,U2:02::}, with estimated sizes for the
possible companions of $\sim$1--4 Jupiter radii.  The large number of
relatively faint ($V =$ 14--18~mag) candidates to study led to our
strategy of a preliminary spectroscopic reconnaissance to detect and
reject large-amplitude (high-mass) companions, followed by more
precise observations of the very best candidates that remained. Of the
59 OGLE candidates, 20 were unsuitable: one is a duplicate entry, 4
have no ephemeris (only one transit was recorded), 8 show obvious
signs in the light curve of a secondary eclipse and/or out-of-eclipse
variations (clear indications of a stellar companion), and 7 were
considered too faint to follow up. We undertook low-resolution
spectroscopy of the other 39 candidates in late June and mid-July 2002
on the Tillinghast 1.5-m telescope at the F.~L.~Whipple Observatory
(Arizona) and the 6.5-m Magellan~I Baade telescope at Las Campanas
Observatory (Chile).  These spectra were used to eliminate stellar
binaries, which can produce shallow, planet-like eclipses due to
blending with light from another star, grazing geometry, or the
combination of a large (early-type) primary and a small stellar
secondary, but are betrayed by large, easily-detected velocity
variations (tens of \kms).  We found 25 of the 39 candidates to be
stellar binaries, and 8 to be of early spectral type.  Only 6
solar-type candidates remained with no detected variations at the few
\kms\ level (G.~Torres et al., in prep.). 

Subsequently, we used the high resolution echelle spectrograph
(HIRES)\cite{Vog:94::} on the Keck~I 10-m telescope at the W.~M.~Keck
Observatory (Hawaii) on the nights of July 24-27, 2002, to obtain
spectra of 5 of these candidates and measure more precise velocities.
OGLE-TR-3 turned out to be the result of grazing eclipses and blending
(with even a hint of a secondary eclipse present in the light curve),
and the data for OGLE-TR-33, OGLE-TR-10, and OGLE-TR-58 are as yet
inconclusive and require further measurements. OGLE-TR-33 exhibits a
complex spectral line profile behaviour and could also be a blend.
OGLE-TR-10 shows insignificant velocity variation, which is consistent
with a sub-Jovian mass planetary companion; OGLE-TR-58 is still
inconclusive because of the uncertain ephemeris (M.~Konacki et al., in
prep.).  Only OGLE-TR-56 showed clear low-amplitude velocity changes
consistent with its 1.21190-day photometric variation\cite{U2:02::},
revealing the planetary nature of the companion.  With only one
bona-fide planet (or at most 3) among the $39+8$ objects examined
spectroscopically or ruled out on the basis of their light curves, the
yield of planets in this particular photometric search has turned out
to be very low: at least 94\% (possibly up to 98\%) of the candidates
are ``false positives".  This is likely to be due in part to the
crowded field towards the Galactic centre, which increases the
incidence of blends. 

We report here our results for OGLE-TR-56 ($I \simeq 15.3$ mag).
Radial velocities were obtained using exposures of a Th-Ar lamp before
and after the stellar exposure for wavelength calibration, and
standard cross-correlation against a carefully-matched synthetic
template spectrum (see Table~1).  Our nightly-averaged measurements
rule out a constant velocity at the 99.3\% confidence level, and are
much better represented by a Keplerian model of an orbiting planet
(Figure~1a and 1b).  Note that the period and phase of the solid curve
are entirely fixed by the transit photometry, as the ephemeris is
constrained extremely well by the 12 transits detected so far
(A.~Udalski, private communication). The only remaining free
parameters are the amplitude of the orbital motion (the key to
establishing the mass of the companion) and the centre-of-mass
velocity, both of which can be accurately determined from our velocity
measurements. The properties of the planet and the star are summarised
in Table~2, and Figure~1c shows a phased light curve of the transit
together with our fitted model.

We performed numerous tests to place limits on any systematic errors
in our radial velocities and to examine other possible causes for the
variation. These are crucial to assess the reality of our detection.
On each night we observed two ``standards" (HD~209458 and HD~179949)
which harbor close-in planets with known
orbits\cite{Mazeh:00::,Tinney:01::}. We derived radial velocities
using the same Th-Ar method as for OGLE-TR-56, and also using the
I$_2$ gas absorption cell to achieve higher accuracy than is possible
for our faint OGLE candidates. In Figure~2 we show that the measured
velocity difference between our two standards (HD~209458 minus
HD~179949) is similar for the Th-Ar and I$_2$ techniques, and more
importantly, that both are consistent with the expected velocity
change. This indicates that we are able to detect real variations at a
level similar to those we see in OGLE-TR-56. 

We can rule out the possibility that OGLE-TR-56 is a giant star
eclipsed by a smaller main-sequence star, both from a test based on
the star's density inferred directly from the transit light
curve\cite{Seager:02::}, and from the very short orbital period.  We
also examined the spectra for sky/solar spectrum contamination from
scattered moonlight; a very small contribution from this source was
removed using TODCOR, a two-dimensional cross-correlation
technique\cite{Zucker:94::}. The separation between the sky lines and
the stellar lines is large enough ($\sim$30 \kms) that the effect on
our derived velocities is very small. 

Blending of the light with other stars is the most serious
concern\cite{Queloz:01::,San:02::} in a crowded field such as toward
the Galactic centre. We have examined the profiles of the stellar
spectral lines for asymmetries and any phase-dependent variations that
can result from blending.  Very little asymmetry is present, and no
correlation with phase is observed. In addition, we performed
numerical simulations to fit the observed light curve assuming
OGLE-TR-56 is blended with a fainter eclipsing binary.  Extensive
tests show that with a photometric precision similar to the OGLE data
($\sigma \simeq$ 0.003--0.015~mag), almost any transit-like light
curve can be reproduced as a blend, and only with spectroscopy can
these cases be recognized. For each trial simulation, the relative
brightness and velocity amplitude of the primary in the eclipsing
binary can be predicted. Although a good fit to the photometry of
OGLE-TR-56 can indeed be obtained for a model with a single star
blended with a fainter system comprising a G star eclipsed by a late M
star, the G star would be bright enough that it would introduce strong
line asymmetries (which are not seen), or would be detected directly
by the presence of a second set of lines in the spectrum.  Careful
inspection using TODCOR\cite{Zucker:94::} rules this out as well.
Therefore, based on the data available, a blend scenario seems
extremely unlikely. 

This is the faintest ($V \simeq 16.6$~mag) and most distant
($\sim1500$~pc) star around which a planet with a known orbit has been
discovered. The planet is quite similar to the only other extrasolar
giant planet with a known radius, HD~209458b, except for having an
orbit which is almost two times smaller. Thus its substellar
hemisphere can heat up to $\sim$1900~K.  However, this is still
insufficient to cause appreciable planet evaporation (with thermal
r.m.s.\ velocity for hydrogen of $\sim$7 \kms\ compared to a surface
escape velocity of $\sim$50 \kms). The tidal Roche lobe radius of
OGLE-TR-56b at its distance from the star is $\sim$2 planet radii.
The planet's orbit is most likely circularised ($e=0.0$) and its
rotation tidally locked, but the star's rotation is not synchronised
($v \sin i \simeq$ 3 \kms). Thus the system appears to have adequate
long-term stability. Interestingly, OGLE-TR-56b is the first planet
found in an orbit much shorter than the current cutoff of close-in
giant planets at 3--4 day periods
($\sim$0.04~AU)\cite{Schneider:02::}.  This might indicate a different
mechanism for halting migration in a protoplanetary disk.  For
example, OGLE-TR-56b may be representative of a very small population
of objects --- the so-called class~II planets, which have lost some of
their mass through Roche lobe overflow\cite{Trilling:98::} but
survived in close proximity to the star; a detailed theoretical study
of OGLE-TR-56b will be presented elsewhere (D.~Sasselov, in prep.).
These observations clearly show that transit searches provide a useful
tool in adding to the amazing diversity of extrasolar planets being
discovered.

\bibliographystyle{nature}

\begin{thebibliography}{10}

\bibitem[{Charbonneau} {\it et~al.}<1>]{Char:00::}
{Charbonneau}, D., {Brown}, T.~M., {Latham}, D.~W. \& {Mayor}, M. {Detection of
  Planetary Transits Across a Sun-like Star}.
\newblock {\it Astrophys. J.} {\bf 529}, L45--L48 (2000).

\bibitem[{Henry} {\it et~al.}<2>]{Hen:00::}
{Henry}, G.~W., {Marcy}, G.~W., {Butler}, R.~P. \& {Vogt}, S.~S. {A Transiting
  ``51 Peg-like'' Planet}.
\newblock {\it Astrophys. J.} {\bf 529}, L41--L44 (2000).

\bibitem[{Mazeh} {\it et~al.}<3>]{Mazeh:00::}
{Mazeh}, T. \emph{et~al.} {The Spectroscopic Orbit of the Planetary Companion
  Transiting HD 209458}.
\newblock {\it Astrophys. J.} {\bf 532}, L55--L58 (2000).

\bibitem[{Udalski} {\it et~al.}<4>]{U1:02::}
{Udalski}, A. \emph{et~al.} {The Optical Gravitational Lensing Experiment.
  Search for Planetary and Low-Luminosity Object Transits in the Galactic Disk.
  Results of 2001 Campaign}.
\newblock {\it Acta Astronomica} {\bf 52}, 1--37 (2002).

\bibitem[{Udalski} {\it et~al.}<5>]{U2:02::}
{Udalski}, A. \emph{et~al.} {The Optical Gravitational Lensing Experiment.
  Search for Planetary and Low- Luminosity Object Transits in the Galactic
  Disk. Results of 2001 Campaign -- Supplement}.
\newblock {\it Acta Astronomica} {\bf 52}, 115--128 (2002).

\bibitem[{Mayor} \& {Queloz}<6>]{May:95::}
{Mayor}, M. \& {Queloz}, D. {A Jupiter-Mass Companion to a Solar-Type Star}.
\newblock {\it Nature} {\bf 378}, 355--359 (1995).

\bibitem[{Marcy} \& {Butler}<7>]{Mar:98::}
{Marcy}, G.~W. \& {Butler}, R.~P. {Detection of Extrasolar Giant Planets}.
\newblock {\it Ann. Rev. Astr. Ap.} {\bf 36}, 57--98 (1998).

\bibitem[{Schneider}<8>]{Schneider:02::}
{Schneider}, J. The Extrasolar Planet Encyclopaedia, web page
  http://www.obspm.fr/encycl/encycl.html (2002).

\bibitem[{Wolszczan} \& {Frail}<9>]{Wol:92::}
{Wolszczan}, A. \& {Frail}, D.~A. {A planetary system around the millisecond
  pulsar PSR1257 + 12}.
\newblock {\it Nature} {\bf 355}, 145--147 (1992).

\bibitem[{Jha} {\it et~al.}<10>]{Jha:00::}
{Jha}, S. \emph{et~al.} {Multicolor Observations of a Planetary Transit of HD
  209458}.
\newblock {\it Astrophys. J.} {\bf 540}, L45--L48 (2000).

\bibitem[{Brown} {\it et~al.}<11>]{Brown:01::}
{Brown}, T.~M., {Charbonneau}, D., {Gilliland}, R.~L., {Noyes}, R.~W. \&
  {Burrows}, A. {Hubble Space Telescope Time-Series Photometry of the
  Transiting Planet of HD 209458}.
\newblock {\it Astrophys. J.} {\bf 552}, 699--709 (2001).

\bibitem[{Charbonneau} {\it et~al.}<12>]{Charbonneau:02::}
{Charbonneau}, D., {Brown}, T.~M., {Noyes}, R.~W. \& {Gilliland}, R.~L.
  {Detection of an Extrasolar Planet Atmosphere}.
\newblock {\it Astrophys. J.} {\bf 568}, 377--384 (2002).

\bibitem[{Brown}, {Libbrecht} \& {Charbonneau}<13>]{Brown:02::}
{Brown}, T.~M., {Libbrecht}, K.~G. \& {Charbonneau}, D. {A Search for CO
  Absorption in the Transmission Spectrum of HD 209458b}.
\newblock {\it Publ. Astr. Soc. Pacific} {\bf 114}, 826--832 (2002).

\bibitem[{Horne}<14>]{Horne:03::}
{Horne}, K. Status and Prospects of Planetary Transit Searches: Hot Jupiters
  Galore, to appear in \emph{Proc. Scientific Frontiers of Research on
  Extrasolar Planets} (eds. Deming, D. \& Seager, S.), in the press (ASP, San
  Francisco, 2003).

\bibitem[{Vogt} {\it et~al.}<15>]{Vog:94::}
{Vogt}, S.~S. \emph{et~al.} in {\it Proc. SPIE Instrumentation in Astronomy
  VIII} (eds Crawford, D.~L. \& Craine, E.~R.)  362--375 (SPIE, Bellingham, WA,
  1994).

\bibitem[{Tinney} {\it et~al.}<16>]{Tinney:01::}
{Tinney}, C.~G. \emph{et~al.} {First Results from the Anglo-Australian Planet
  Search: A Brown Dwarf Candidate and a 51 Peg-like Planet}.
\newblock {\it Astrophys. J.} {\bf 551}, 507--511 (2001).

\bibitem[{Seager} \& {Mall\'en-Ornelas}<17>]{Seager:02::}
{Seager}, S. \& {Mall\'en-Ornelas}, G. On the Unique Solution of Planet and
  Star Parameters from an Extrasolar Planet Transit Light Curve.
  \emph{Astrophys. J.} (in press); preprint http://arXiv.org/astro-ph/0206228
  (2002).

\bibitem[{Zucker} \& {Mazeh}<18>]{Zucker:94::}
{Zucker}, S. \& {Mazeh}, T. {Study of spectroscopic binaries with TODCOR. I: A
  new two-dimensional correlation algorithm to derive the radial velocities of
  the two components}.
\newblock {\it Astrophys. J.} {\bf 420}, 806--810 (1994).

\bibitem[{Queloz} {\it et~al.}<19>]{Queloz:01::}
{Queloz}, D. \emph{et~al.} {No planet for HD 166435}.
\newblock {\it Astr. Astrophys.} {\bf 379}, 279--287 (2001).

\bibitem[{Santos} {\it et~al.}<20>]{San:02::}
{Santos}, N.~C. \emph{et~al.} {The CORALIE survey for southern extra-solar
  planets. IX. A 1.3-day period brown dwarf disguised as a planet}.
\newblock {\it Astr. Astrophys.} {\bf 392}, 215--229 (2002).

\bibitem[{Trilling} {\it et~al.}<21>]{Trilling:98::}
{Trilling}, D.~E. \emph{et~al.} {Orbital Evolution and Migration of Giant
  Planets: Modeling Extrasolar Planets}.
\newblock {\it Astrophys. J.} {\bf 500}, 428--439 (1998).

\bibitem[{Barlow}<22>]{Barlow:01::}
{Barlow}, T. MAKEE Keck Observatory HIRES Data Reduction Software, web page
  http://spider.ipac.caltech.edu/staff/tab/makee/ (2001).

\bibitem[{Cody} \& {Sasselov}<23>]{Cody:02::}
{Cody}, A.~M. \& {Sasselov}, D.~D. {HD 209458: Physical Parameters of the
  Parent Star and the Transiting Planet}.
\newblock {\it Astrophys. J.} {\bf 569}, 451--458 (2002).

\end{thebibliography}

\section*{Acknowledgements} 

We wish to thank A.~Udalski and the OGLE team for generous
contributions to this project.  We are very grateful to S.~Kulkarni
for invaluable support, to R.~Noyes and D.~Latham for helpful
comments, to T.~Barlow for assistance with the spectroscopic
reductions using MAKEE\cite{Barlow:01::}, and to K.~Stanek for his
continuous encouragement.  The data presented herein were obtained at
the W.~M.~Keck Observatory (operated by Caltech, Univ.~of California,
and NASA), which was made possible by the generous financial support
of the W.~M.~Keck Foundation. M.K.\ gratefully acknowledges the
support of NASA through the Michelson Fellowship
programme. G.T. acknowledges support from NASA's Kepler Mission. S.J.\
thanks the Miller Institute for Basic Research in Science at UC
Berkeley for support via a research fellowship.

Correspondence and requests for materials should be addressed to M.K.
(e-mail: maciej@gps.caltech.edu)

\clearpage

\begin{table}[!t]
\begin{center}
\begin{tabular}{|c|c|c|}
\hline
Date        & RV       & Error \\
MJD         & \kms     & \kms  \\
\hline\hline
52480.4239  & $-$49.26   & 0.20 \\
52481.4011  & $-$49.44   & 0.08 \\
52481.4178  & $-$49.24   & 0.09 \\
52483.3984  & $-$49.60   & 0.06 \\
52483.4152  & $-$49.78   & 0.11 \\
\hline
\end{tabular}
\end{center}
\caption[]{{\bf OGLE-TR-56 radial velocities.} The velocities (reduced
to the solar system barycentre) and formal errors are given for each
of our individual spectra of OGLE-TR-56.  The data indicate a
significant variation; a flat line fit gives $\chi^2 \simeq 20$ with 4
degrees of freedom (0.06\% false alarm probability), which is
considerably worse than a fit to a Keplerian orbit model with a fixed
ephemeris ($\chi^2 \simeq 5$ with 3 degrees of freedom; 17\%
probability). Having shown, as a check, that the velocities from
separate exposures on the same night (originally intended for cosmic
ray removal) are not significantly different, we have adopted the
nightly averages for subsequent use. A similarly high significance is
found for the conclusion that the average velocities are not well fit
by a flat line (99.3\% confidence level).}
\end{table}

\begin{table}[!b]
\begin{center}
\begin{tabular}{|lc|}
\hline
Parameter        & Value \\
\hline\hline
Velocity amplitude          & $0.167 \pm 0.027$ \kms \\
Centre-of-mass velocity     & $-49.49 \pm 0.02$ \kms \\
Orbital period              & $1.21190 \pm 0.00001$ days \\
Reference transit epoch (MJD)    & $52072.185 \pm 0.003$ \\
Star mass                   & $1.04 \pm 0.05$ M$_{\odot}$ \\
Star radius                 & $1.10 \pm 0.10$ R$_{\odot}$ \\
Limb darkening coefficient ($I$ band)     & $0.56 \pm 0.06$\\
Orbital inclination         & $86 \pm 2$ deg \\
Planet distance from star   & 0.0225 AU \\
Planet mass                 & $0.9 \pm 0.3$ M$_{\rm Jup}$ \\
Planet radius               & $1.30 \pm 0.15$ R$_{\rm Jup}$  \\
Planet density              & $0.5 \pm 0.3$ g~cm$^{-3}$ \\
\hline
\end{tabular}
\end{center}
 \caption[]{{\bf Derived stellar and planetary parameters.} The physical
properties of the star were derived by modelling the high-resolution
spectra with numerical model atmospheres. We find that OGLE-TR-56 is
very similar to the Sun, with a temperature of $T_{\rm eff} \sim
5900$~K.  The star's mass and radius were computed from our stellar
evolution tracks\cite{Cody:02::}. Combining the stellar parameters
with the OGLE-III $I$-band photometry yields the planetary radius and
orbital inclination.  The uncertainties shown for the orbital elements
are formal errors; the errors for the planet mass and radius
additionally reflect our conservative estimate of systematic
uncertainties.} \end{table}

\clearpage

\begin{figure}[!h] 
\centerline{{\psfig{file=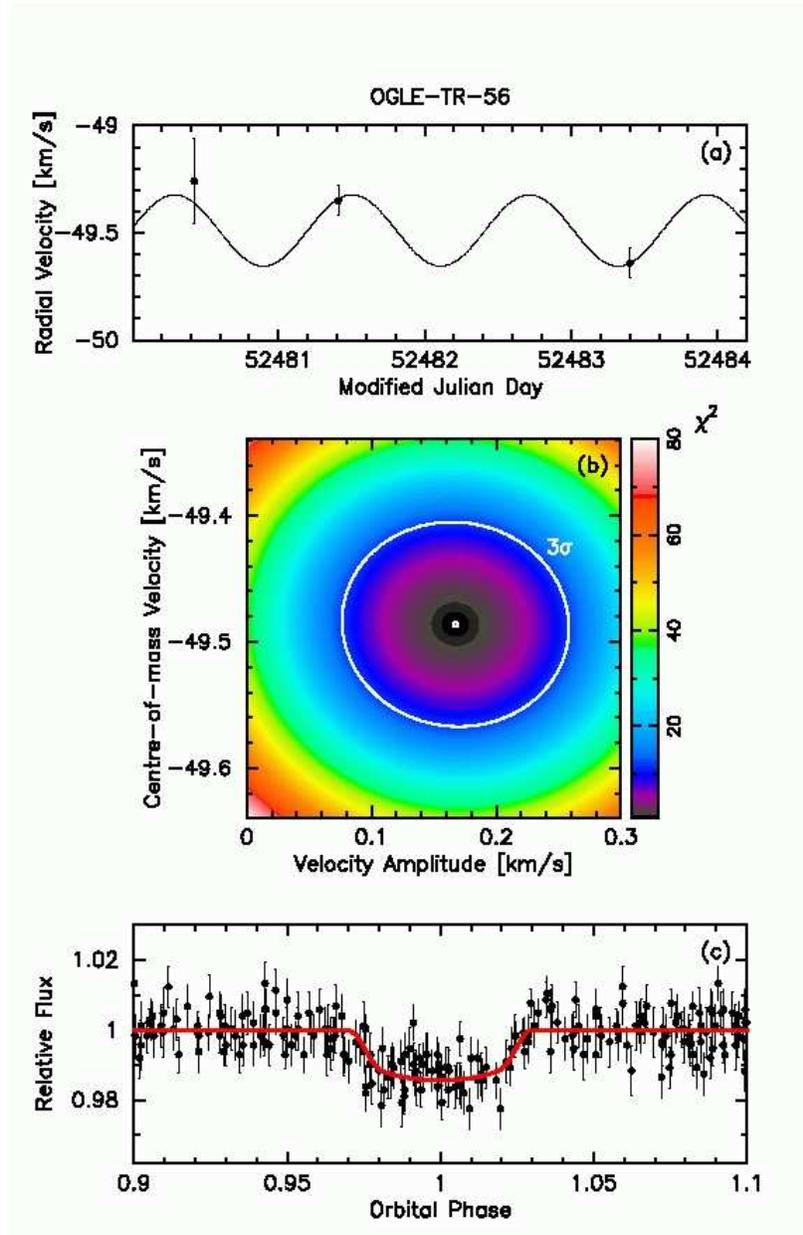,height=6.5in}}}
 \caption[]{{\bf Spectroscopic and photometric
observations.} (a) Our radial velocity measurements for OGLE-TR-56
(nightly averages).  Note the good agreement with a curve whose only
free parameters are the amplitude and systemic velocity; the period
and transit epoch are fixed from the OGLE-III photometry. The curve
assumes a circularised orbit ($e=0.0$), as is theoretically expected,
and the line thickness corresponds to the phase uncertainty.  (b)
$\chi^2$ surface showing the confidence region for our determination
of the velocity amplitude and centre-of-mass velocity.  The detection
of a change in velocity (nonzero velocity amplitude) is formally
significant at 6$\sigma$ (see also Table~2).  (c) The photometric
transit light curve of OGLE-TR-56 from Udalski et al.\cite{U2:02::}
The transit has an extended flat bottom, and its 1.2\%-depth points to
a Jupiter-size body (given our determination of a Sun-like primary).
The solid line represents our fitted solution to derive the system
parameters. The light curve shows no evidence of other variations or
of a secondary eclipse (the hallmark of a strongly blended stellar
binary).} \end{figure}

\begin{figure}[!h] 
 \centerline{\psfig{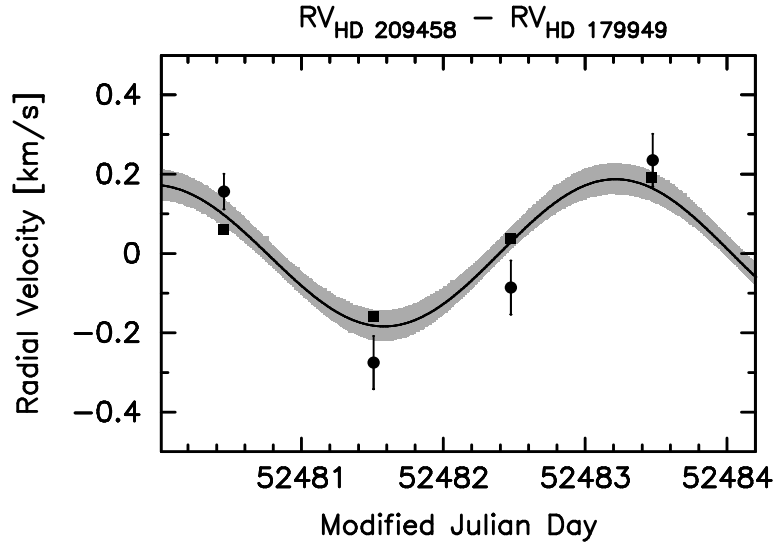}}
 \caption[]{{\bf Tests for systematic errors.} Predicted radial
velocity {\em difference} between our two standard stars with known
planets --- HD~209458\cite{Mazeh:00::} and HD~179949\cite{Tinney:01::}
(solid line), compared with measurements on each of our four observing
nights in July 2002. The filled circles are our Th-Ar velocity
differences (HD~209458 minus HD~179949, from the blue echelle orders
beyond the iodine spectrum cutoff) with a typical internal uncertainty
of about 100 \mps. These differences should be independent of the
wavelength solution itself, and should reveal only the real difference
in the Doppler shifts of the stars as well as any systematic problems
of an instrumental nature.  For comparison, our more precise
iodine-cell velocity differences for the same stars (squares) have
typical uncertainties of 20 \mps.  The uncertainty introduced by
errors in the orbital elements of the standards is indicated by the
shaded area.  The graph shows that we have succeeded in measuring
small changes in velocity on different nights using the standard Th-Ar
technique, which reflects on the excellent stability of the HIRES
instrument. The same technique was applied to our observations of
OGLE-TR-56.}
 \end{figure}

\end{document}